\documentclass[preprintnumbers]{revtex4}
\usepackage{amsmath}
\usepackage{amssymb}
\usepackage{graphicx}
\usepackage{color}

\def\bc{\begin{center}}
\def\ec{\end{center}}

\def\beq{\begin{equation}}
\def\eeq{\end{equation}}

\begin{document}

\title{Charge-dipole and dipole-dipole interactions in two-dimensional
materials}
\author{ Roman Ya. Kezerashvili$^{1,2}$ and Vladimir Ya. Kezerashvili$^{1,3}$
}
\affiliation{\mbox{$^{1}$Physics Department, New York
City College
of Technology, The City University of New York,} \\
Brooklyn, NY 11201, USA \\
\mbox{$^{2}$The Graduate School and University Center, The
City University of New York,} \\
New York, NY 10016, USA \\
\mbox{$^{3}$Borough of Manhattan Community College, The
City University of New York,} \\
New York, NY 10007, USA \\
\\
}
\date{\today}

\begin{abstract}
We derive the explicit analytical form for the charge-dipole and
dipole-dipole interactions in 2D configuration space. We demonstrate that
the reduction of dimensionality can alter the charge-dipole and
dipole-dipole interactions in 2D case. The asymptotics of these interactions
at large distances coincide to the charge-dipole and dipole-dipole
interactions in 3D configuration space.

{\ }
\end{abstract}

\maketitle



\label{intro}

In classical electrodynamics for description of the field produced by a
system of electric charges at large distances the concepts of dipoles and
multipole moments are very important and well developed \cite%
{Landau2,Jackson}. This approach is based on the potential of a single
charge in three-dimensional (3D) configuration space. Ordinary matter is
more or less uncharged, but it is reach in pair of charges called dipoles.
Dipoles are building blocks of bulk dielectric and magnetic materials. Not
surprisingly, it turns out to be efficient mathematically to deal with the
dipole not as just a pair of individual positive and negative charges.

The last two decades discoveries and studies of two-dimensional (2D)
materials attract a considerable interest. Atomically thin materials such as
graphene and monolayer transition metal dichalcogenides (TMDC), phosphorene,
Xenes (silicene, germanine, stanene), exhibit remarkable physical properties
resulting from their reduced dimensionality and crystal symmetry. The family
of semiconducting transition metal dichalcogenides is an especially
promising platform for fundamental studies of two-dimensional systems, with
potential applications in optoelectronics and valleytronics due to their
direct band gap in the monolayer. Exciton is the simplest bound complex
formed by an electron in a conduction band and hole in a valence band. The
description of excitons, trions, beexcitons in 2D material requires
knowledge of electrostatic interaction in reduced dimensionality.

The interaction of two charge particles in two-dimensional space are studied
in detail and the analytical expression for two charged particle interaction
is well known \cite{Rytova1967, Keldysh1979} and widely used for description
of excitonic complexes in 2D materials (see reviews: \cite%
{Kormanyos,Glazov,Kezerashvili}). In 3D configuration space when charged
particles interact via the Coulomb potential the corresponding charge-dipole
and dipole-dipole potentials are well known. In contrast, the influence of
the reduction of dimensionality on the charge-dipole and dipole-dipole
interactions in 2D configuration space has not yet been investigated. We
still lack of the analytical expression for the charge-dipole and
dipole-dipole 2D interactions. Below we derive the explicit analytical form
for the charge-dipole and dipole-dipole interactions in 2D space.\label{C_C}

\textit{Charge-charge interaction in 2D configuration space.} An interaction
of two charged particles in the context of thin semiconductor films, was
derived analytically by Rytova \cite{Rytova1967} and,\ decade later, by
Keldysh \cite{Keldysh1979}. Due to the lack of screening by the environment
above the material layer it was shown that the electron-hole interaction
potential in a thin semiconductor layer is not Coulombic. Over the course of
decade the celebrated Rytova-Keldysh (RK) potential \cite%
{Rytova1967,Keldysh1979} has been widely used to describe the electrostatic
interaction of few-body complexes such as excitons, trions, and biexcitons
\cite{Kezerashvili} in monolayer transition-metal dichalcogenides,
phosphorene and Xenes. This potential describes the non-hydrogenic Rydberg
series of neutral excitons. The effective electron-hole Rytova-Keldysh
potential, which takes into account screening due to the reduction of
dimensionality is given by \cite{Rytova1967,Keldysh1979}:

\begin{equation}
V_{\text{RK}}\left( R\right) =-\frac{\pi ke^{2}}{2\kappa \rho _{0}}\left[
H_{0}\left( \frac{R}{\rho _{0}}\right) -Y_{0}\left( \frac{R}{\rho _{0}}%
\right) \right] .  \label{eq:vkeld}
\end{equation}%
In Eq.~\eqref{eq:vkeld}, $k=9\times 10^{9}$ N$\cdot $m$^{2}$/C$^{2}$, $%
R=\left\vert \mathbf{R}\right\vert $ is the magnitude of the relative
electron-hole separation, $\kappa =(\epsilon _{1}+\epsilon _{2})/2$
describes the surrounding dielectric environment, where $\epsilon _{1}$ and $%
\epsilon _{2}$ correspond to the dielectric constants of the materials above
and below the monolayer, $H_{0}$ and $Y_{0}$ are the Struve and Bessel
functions of the second kind, respectively, and $\rho _{0}$ is the screening
length. In the case of a thin semiconductor layer of a finite thickness $t$
and an isotropic dielectric constant $\varepsilon $, the screening length is
evaluated as\ $\rho _{0}=t\varepsilon /2\kappa $\ \cite{Keldysh1979}. In the
case of atomically thin 2D materials the screening length is given by~\cite%
{Cudazzo2011,Berkelbach2013}:
\begin{equation}
\rho _{0}=\frac{2\pi \chi _{2D}}{\kappa },  \label{eq:rho0chi2d}
\end{equation}%
where $\chi _{2D}$ is the 2D polarizability, which can be calculated via
\textit{ab-initio} methods or considered as a phenomenological parameter.
The screening length typically ranges from roughly 30 to 80 \AA\ \cite%
{Glazov}. The effective interaction potential \eqref{eq:vkeld} has an
asymptotic behavior $\sim 1/R$ only at large distances between the
particles, that follows from \cite{Gradshteyn,Abramowitz}. This limiting
case corresponds to the Coulomb interaction unaffected by the dielectric
polarization of a 2D layer, as most of the electric-field lines between two
distant charges go outside of the 2D semiconductor. Interestingly enough, in
this limiting case two charges are interacting the same way as in vacuum. At
smaller distances the potential deviates strongly from the usual $1/R$ form
and the dependence has a logarithmic behavior $\sim \ln (2/r)-\gamma $ \cite%
{Gradshteyn,Abramowitz,Cudazzo2011}, where $\gamma =0.5772...$ is Euler's
constant.
\begin{figure}[t]
\centering
\includegraphics[width=12.cm]{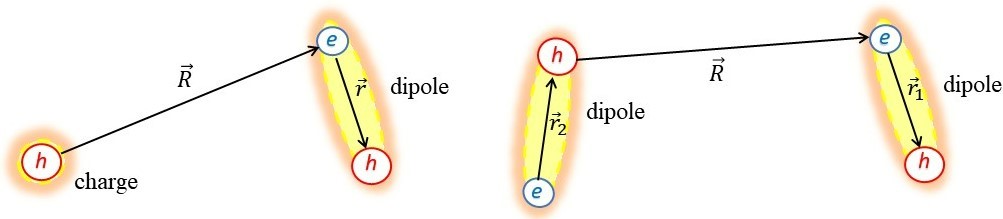}
\caption{(Color online) Schematics for the charge-dipole and dipole-dipole
interactions in 2D configuration space. }
\label{CD}
\end{figure}

\label{C_D}

\textit{Charge-dipole interaction in 2D configuration space.} Consider two
opposite closely spaced charges in a monolayer that form a dipole with the
dipole moment $\mathbf{d}=e\mathbf{r}$ and a single charge placed at a
distance $R$ as shown in Fig. \ref{CD}. The point charge interacts with the
dipole via the RK potential. In this case following notations in Fig. \ref%
{CD} for the charge-dipole interaction we have

\begin{equation}
V_{\text{cd}}\left( \mathbf{R}\right) \equiv V_{\text{eh}}\left( R\right)
+V_{\text{hh}}\left( \left\vert \mathbf{R}+\mathbf{r}\right\vert \right) =-%
\frac{\pi ke^{2}}{2\kappa \rho _{0}}\left[ H_{0}\left( \frac{R}{\rho _{0}}%
\right) -Y_{0}\left( \frac{R}{\rho _{0}}\right) \right] +\frac{\pi ke^{2}}{%
2\kappa \rho _{0}}\left[ H_{0}\left( \frac{\left\vert \mathbf{R}+\mathbf{r}%
\right\vert }{\rho _{0}}\right) -Y_{0}\left( \frac{\left\vert \mathbf{R}+%
\mathbf{r}\right\vert }{\rho _{0}}\right) \right] ,  \label{CD1}
\end{equation}%
where
\begin{equation}
\left\vert \mathbf{R}+\mathbf{r}\right\vert =R\sqrt{1+\frac{2\mathbf{R\cdot r%
}}{R^{2}}+\frac{r^{2}}{R^{2}}}.
\end{equation}%
For $R\gg r$ $\left( 1+\frac{2\mathbf{R\cdot r}}{R^{2}}+\frac{r^{2}}{R^{2}}%
\right) ^{1/2}\simeq 1+\frac{1}{2}\left( \frac{2\mathbf{R\cdot r}}{R^{2}}+%
\frac{r^{2}}{R^{2}}\right) $. Considering only linear terms with respect to $%
r$, Eq. (\ref{CD1}) can be written as

\begin{equation}
V_{\text{cd}}\left( \mathbf{R}\right) =-\frac{\pi ke^{2}}{2\kappa \rho _{0}}%
\left[ H_{0}\left( \frac{R}{\rho _{0}}\right) -Y_{0}\left( \frac{R}{\rho _{0}%
}\right) \right] +\frac{\pi ke^{2}}{2\kappa \rho _{0}}\left[ H_{0}\left(
\frac{R}{\rho _{0}}\left[ 1+\frac{\mathbf{R\cdot r}}{R^{2}}\right] \right)
-Y_{0}\left( \frac{R}{\rho _{0}}\left[ 1+\frac{\mathbf{R\cdot r}}{R^{2}}%
\right] \right) \right] .  \label{CD0}
\end{equation}%
Expand the Struve $H_{0}\left( \frac{R}{\rho _{0}}\left[ 1+\frac{\mathbf{%
R\cdot r}}{R^{2}}\right] \right) $ and Bessel $Y_{0}\left( \frac{R}{\rho _{0}%
}\left[ 1+\frac{\mathbf{R\cdot r}}{R^{2}}\right] \right) $ functions in
terms of power of $\frac{r}{R}$ when $R\gg r$ and consider linear terms with
respect to $r$:
\begin{eqnarray}
H_{0}\left( \frac{R}{\rho _{0}}\left[ 1+\frac{\mathbf{R\cdot r}}{R^{2}}%
\right] \right) &\simeq &H_{0}\left( \frac{R}{\rho _{0}}\right) +\left.
H_{0}^{^{\prime }}\left( x\right) \right\vert _{x=\frac{R}{\rho _{0}}}\frac{R%
}{\rho _{0}}\frac{\mathbf{R\cdot r}}{R^{2}}=H_{0}\left( \frac{R}{\rho _{0}}%
\right) +H_{-1}\left( \frac{R}{\rho _{0}}\right) \frac{R}{\rho _{0}}\frac{%
\mathbf{R\cdot r}}{R^{2}},  \label{SF1} \\
Y_{0}\left( \frac{R}{\rho _{0}}\left[ 1+\frac{\mathbf{R\cdot r}}{R^{2}}%
\right] \right) &\simeq &Y_{0}\left( \frac{R}{\rho _{0}}\right) +\left.
Y_{0}^{^{\prime }}\left( x\right) \right\vert _{x=\frac{R}{\rho _{0}}}\frac{R%
}{\rho _{0}}\frac{\mathbf{R\cdot r}}{R^{2}}=Y_{0}\left( \frac{R}{\rho _{0}}%
\right) +Y_{-1}\left( \frac{R}{\rho _{0}}\right) \frac{R}{\rho _{0}}\frac{%
\mathbf{R\cdot r}}{R^{2}}.  \label{BF1}
\end{eqnarray}%
Here we use that $H_{0}^{^{\prime }}(x)=H_{-1}\left( x\right) $ and $%
Y_{0}^{^{\prime }}(x)=-Y_{1}(x)=Y_{-1}\left( x\right) $ \cite%
{Gradshteyn,Abramowitz,Olver}. Using (\ref{SF1}) and (\ref{BF1}) Eq. (\ref%
{CD0}) can be written as
\begin{equation}
V_{\text{cd}}\left( \mathbf{R}\right) =\frac{\pi ke^{2}}{2\kappa \rho
_{0}^{2}}\left[ H_{-1}\left( \frac{R}{\rho _{0}}\right) -Y_{-1}\left( \frac{R%
}{\rho _{0}}\right) \right] \frac{\mathbf{R\cdot r}}{R},  \label{CDKel}
\end{equation}%
or
\begin{equation}
V_{\text{cd}}\left( \mathbf{R}\right) =\frac{\pi ke}{2\kappa \rho _{0}}\left[
H_{-1}\left( \frac{R}{\rho _{0}}\right) -Y_{-1}\left( \frac{R}{\rho _{0}}%
\right) \right] \frac{\mathbf{R\cdot d}}{\rho _{0}R}.  \label{cdr}
\end{equation}

Consider the asymptotic of $V_{\text{cd}}\left( \mathbf{R}\right) $
interaction when $R\rightarrow \infty .$ For the difference of $H_{-1}\left(
x\right) -Y_{-1}\left( x\right) $ when $x\rightarrow \infty $ we have \cite%
{Gradshteyn,Abramowitz,Paris}

\begin{equation}
H_{\nu }\left( x\right) -Y_{\nu }\left( x\right) \xrightarrow[x\rightarrow
\infty]{}\frac{\left( \frac{x}{2}\right) ^{\nu -1}}{\sqrt{\pi }\Gamma (\nu
+1/2)}.  \label{HYAsymp}
\end{equation}%
For $\nu =-1$ $H_{-1}\left( \frac{R}{\rho _{0}}\right) -Y_{-1}\left( \frac{R%
}{\rho _{0}}\right) =\frac{4\rho _{0}^{2}}{R^{2}}\frac{1}{\sqrt{\pi }\Gamma
(-1/2)}$. Therefore,

\begin{equation}
V_{\text{cd}}\left( \mathbf{R}\right) \xrightarrow[R \rightarrow \infty]{}-%
\frac{ke}{\kappa }\frac{\mathbf{R\cdot d}}{R^{3}},  \label{Asymptotic}
\end{equation}%
where we use $\Gamma (-1/2)=-2\sqrt{\pi }.$ Thus, one can conclude that in
2D configuration space the charge-dipole interaction has the form (\ref{cdr}%
) that has the asymptotic (\ref{Asymptotic}).

Evidently, the charge-dipole potential goes like $\frac{1}{R^{2}}$ at large $%
R$ and it falls off more rapidly than potential \eqref{eq:vkeld}. For the
charge-dipole interaction in 3D configuration space we have well known
expression

\begin{equation}
V_{\text{cd}}^{C}\left( \mathbf{R}\right) =-\frac{ke}{\kappa }\frac{\mathbf{%
R\cdot d}}{R^{3}}.\text{ }  \label{CDcoul}
\end{equation}%
We can conclude that\ $V_{\text{cd}}\left( \mathbf{R}\right) $ in 2D and $V_{%
\text{cd}}^{C}\left( \mathbf{R}\right) $ in 3D configuration spaces,
respectively, vary as $\frac{1}{R^{2}}$ at large separation of the charge
and the dipole.

\label{D_D}

\textit{Dipole-dipole interaction in 2D configuration space.} Consider two
dipoles $\mathbf{d}_{1}=e\mathbf{r}_{1}$ and $\mathbf{d}_{2}=e\mathbf{r}_{2}$
interaction in 2D configuration space. One can consider a dipole-dipole
interaction as the interactions of a positive and negative charges of one
dipole with the second dipole. Following notations in Fig. \ref{CD} for a
dipole-dipole interaction in 2D space we have

\begin{equation}
V_{\text{dd}}\left( \mathbf{R}\right) =V_{\text{hd}}\left( \frac{R}{\rho _{0}%
}\right) +V_{\text{ed}}\left( \frac{\left\vert \mathbf{R}+\mathbf{r}%
_{2}\right\vert }{\rho _{0}}\right) .  \label{DD}
\end{equation}%
Using (\ref{cdr}) for the charge-dipole interaction in Eq. (\ref{DD}), we
have
\begin{equation}
V_{\text{dd}}\left( \mathbf{R}\right) =\frac{\pi ke^{2}}{2\kappa \rho
_{0}^{2}}\left[ H_{-1}\left( \frac{R}{\rho _{0}}\right) -Y_{-1}\left( \frac{R%
}{\rho _{0}}\right) \right] \frac{\mathbf{R\cdot r}_{1}}{R}-\frac{\pi ke}{%
2\kappa \rho _{0}^{2}}\left[ H_{-1}\left( \frac{\left\vert \mathbf{R+r}%
_{2}\right\vert }{\rho _{0}}\right) -Y_{-1}\left( \frac{\left\vert \mathbf{%
R+r}_{2}\right\vert }{\rho _{0}}\right) \right] \frac{\left( \mathbf{R+r}%
_{2}\right) \mathbf{\cdot r}_{1}}{\left\vert \mathbf{R+r}_{2}\right\vert }.
\label{ddr1r2}
\end{equation}%
We focus on the second term in Eq. (\ref{DD}). For $R\gg r_{2}$ considering
the terms linear with respect to $r_{2}$, we have $\left\vert \mathbf{R+r}%
_{2}\right\vert =R\left( 1+\frac{2\mathbf{R\cdot r}_{2}}{R^{2}}+\frac{%
r_{2}^{2}}{R^{2}}\right) ^{1/2}\simeq R(1+\frac{\mathbf{R\cdot r}_{2}}{R^{2}}%
)$ and $\frac{1}{\left\vert \mathbf{R+r}_{2}\right\vert }=\frac{1}{R}\left(
1+\frac{2\mathbf{R\cdot r}_{2}}{R^{2}}+\frac{r_{2}^{2}}{R^{2}}\right)
^{-1/2}\simeq \frac{1}{R}\left( 1-\frac{\mathbf{R\cdot r}_{2}}{R^{2}}\right)
$ and (\ref{ddr1r2}) becomes

\begin{eqnarray}
V_{\text{ed}}\left( \frac{\left\vert \mathbf{R}+\mathbf{r}_{2}\right\vert }{%
\rho _{0}}\right) &=&-\frac{\pi ke^{2}}{2\kappa \rho _{0}^{2}}\left[
H_{-1}\left( \frac{\left\vert \mathbf{R+r}_{2}\right\vert }{\rho _{0}}%
\right) -Y_{-1}\left( \frac{\left\vert \mathbf{R+r}_{2}\right\vert }{\rho
_{0}}\right) \right] \frac{\left( \mathbf{R+r}_{2}\right) \mathbf{\cdot r}%
_{1}}{\left\vert \mathbf{R+r}_{2}\right\vert }=  \notag \\
&&-\frac{\pi ke^{2}}{2\kappa \rho _{0}^{2}}\left\{ H_{-1}\left( \frac{R}{%
\rho _{0}}\left[ 1+\frac{\mathbf{R\cdot r}_{2}}{R^{2}}\right] \right)
-Y_{-1}\left( \frac{R}{\rho _{0}}\left[ 1+\frac{\mathbf{R\cdot r}_{2}}{R^{2}}%
\right] \right) \right\} \left( \frac{\left( \mathbf{R+r}_{2}\right) \mathbf{%
\cdot r}_{1}}{R}-\frac{\mathbf{R\mathbf{\cdot r}_{1}\mathbf{R\cdot r}_{2}}}{%
R^{3}}\right) .  \label{eD1}
\end{eqnarray}

Expand the Struve $H_{-1}\left( \frac{R}{\rho _{0}}\left[ 1+\frac{\mathbf{%
R\cdot r}_{2}}{R^{2}}\right] \right) $ and Bessel $Y_{-1}\left( \frac{R}{%
\rho _{0}}\left[ 1+\frac{\mathbf{R\cdot r}_{2}}{R^{2}}\right] \right) $
functions in terms of power of $\frac{r_{2}}{R}$ when $R\gg r_{2}$ and
consider linear terms with respect to $r_{2}$:
\begin{eqnarray}
H_{-1}\left( \frac{R}{\rho _{0}}\left[ 1+\frac{\mathbf{R\cdot r}_{2}}{R^{2}}%
\right] \right) &\simeq &H_{-1}\left( \frac{R}{\rho _{0}}\right) +\left.
H_{-1}^{^{\prime }}\left( x\right) \right\vert _{x=\frac{R}{\rho _{0}}}\frac{%
R}{\rho _{0}}\frac{\mathbf{R\cdot r}_{2}}{R^{2}},  \label{SF2} \\
Y_{-1}\left( \frac{R}{\rho _{0}}\left[ 1+\frac{\mathbf{R\cdot r}_{2}}{R^{2}}%
\right] \right) &\simeq &Y_{-1}\left( \frac{R}{\rho _{0}}\right) +\left.
Y_{-1}^{^{\prime }}\left( x\right) \right\vert _{x=\frac{R}{\rho _{0}}}\frac{%
R}{\rho _{0}}\frac{\mathbf{R\cdot r}_{2}}{R^{2}}.  \label{BF2}
\end{eqnarray}%
When $R\gg r_{1}$ and $R\gg r_{2}$ by considering only terms linear with
respect to $r_{1}$ and $r_{2}$ and using (\ref{SF2}) and (\ref{BF2}) finally
Eq. (\ref{eD1}) can be written as
\begin{eqnarray}
V_{\text{ed}}\left( \frac{\left\vert \mathbf{R}+\mathbf{r}_{2}\right\vert }{%
\rho _{0}}\right) &=&-\frac{\pi ke^{2}}{2\kappa \rho _{0}^{2}}\left[
H_{-1}\left( \frac{R}{\rho _{0}}\right) -Y_{-1}\left( \frac{R}{\rho _{0}}%
\right) \right] \left( \frac{\mathbf{R\mathbf{\cdot r}_{1}}}{R}+\frac{%
\mathbf{\mathbf{r}_{1}\mathbf{\cdot r}_{2}}}{R}-\frac{\mathbf{R\mathbf{\cdot
r}_{1}\mathbf{R\cdot r}_{2}}}{R^{3}}\right) -  \notag \\
&&\frac{\pi ke^{2}}{2\kappa \rho _{0}^{2}}\left[ H_{-1}^{^{\prime }}\left(
\frac{R}{\rho _{0}}\right) -Y_{-1}^{^{\prime }}\left( \frac{R}{\rho _{0}}%
\right) \right] \frac{\mathbf{R\mathbf{\cdot r}_{1}R\cdot r}_{2}}{\rho
_{0}R^{2}}.  \label{eD2}
\end{eqnarray}%
Let us find $H_{-1}^{^{\prime }}\left( \frac{R}{\rho _{0}}\right)
-Y_{-1}^{^{\prime }}\left( \frac{R}{\rho _{0}}\right) $ which presents in (%
\ref{eD2}). The recurrence relations for the Struve functions $H_{\nu
-1}(x)+H_{\nu +1}(x)=\frac{2\nu }{x}H_{\nu }(x)+\left( \frac{x}{2}\right)
^{\nu }\frac{1}{\sqrt{\pi }\Gamma (\nu +3/2)}$ and $H_{\nu -1}(x)-H_{\nu
+1}(x)=2H_{\nu }^{^{\prime }}(x)-\left( \frac{x}{2}\right) ^{\nu }\frac{1}{%
\sqrt{\pi }\Gamma (\nu +3/2)}$ \cite{Gradshteyn,Abramowitz} lead to $H_{\nu }%
{^{\prime }\left( x\right) =H_{\nu -1}\left( x\right) +\frac{1}{x}%
H_{-1}\left( x\right) }$. On the other hand, for the second kind Bessel
function $Y_{\nu }^{^{\prime }}(x)=Y_{\nu -1}(x)-\frac{\nu }{x}Y_{\nu }(x)$
\cite{Gradshteyn}. Therefore, for $\nu =-1$ we obtain: $H_{-1}^{^{\prime
}}\left( \frac{R}{\rho _{0}}\right) -Y_{-1}^{^{\prime }}\left( \frac{R}{\rho
_{0}}\right) =H_{-2}\left( \frac{R}{\rho _{0}}\right) -Y_{-2}\left( \frac{R}{%
\rho _{0}}\right) +\frac{\rho _{0}}{R}\left[ H_{-1}\left( \frac{R}{\rho _{0}}%
\right) -Y_{-1}\left( \frac{R}{\rho _{0}}\right) \right] $. The latter
expression allows rewrite Eq. (\ref{eD2}) as
\begin{eqnarray}
V_{\text{ed}}\left( \frac{\left\vert \mathbf{R}+\mathbf{r}_{2}\right\vert }{%
\rho _{0}}\right) &=&-\frac{\pi ke^{2}}{2\kappa \rho _{0}^{2}}\left[
H_{-1}\left( \frac{R}{\rho _{0}}\right) -Y_{-1}\left( \frac{R}{\rho _{0}}%
\right) \right] \left( \frac{\mathbf{R\mathbf{\cdot r}_{1}}}{R}+\frac{%
\mathbf{\mathbf{r}_{1}\mathbf{\cdot r}_{2}}}{R}\right) -  \notag \\
&&\frac{\pi ke^{2}}{2\kappa \rho _{0}^{2}}\left[ H_{-2}\left( \frac{R}{\rho
_{0}}\right) -Y_{-2}\left( \frac{R}{\rho _{0}}\right) \right] \frac{\mathbf{R%
\mathbf{\cdot r}_{1}R\cdot r}_{2}}{\rho _{0}R^{2}}.  \label{eD3}
\end{eqnarray}%
Replacing the second term in Eq. (\ref{ddr1r2}) by expression (\ref{eD3}) we
obtain the dipole-dipole interaction in 2D configuration space
\begin{equation}
V_{\text{dd}}\left( R\right) =-\frac{\pi k}{2\kappa \rho _{0}}\left\{ \left[
H_{-1}\left( \frac{R}{\rho _{0}}\right) -Y_{-1}\left( \frac{R}{\rho _{0}}%
\right) \right] \frac{\mathbf{\mathbf{d}_{1}\mathbf{\cdot }d_{2}}}{\rho _{0}R%
}+\left[ H_{-2}\left( \frac{R}{\rho _{0}}\right) -Y_{-2}\left( \frac{R}{\rho
_{0}}\right) \right] \frac{\mathbf{R\mathbf{\cdot }d_{1}R\cdot d}_{2}}{\rho
_{0}^{2}R^{2}}\right\} .  \label{DDK}
\end{equation}%
Using Eq. (\ref{HYAsymp}) for $\nu =-1$ and $\nu =-2$ for the first and
second term in Eq. (\ref{DDK}), respectively, one can find the asymptotic of
$V_{\text{dd}}\left( \mathbf{R}\right) $ interaction when $R\rightarrow
\infty .$ The first term in Eq. (\ref{DDK}) has the following asymptotic
behavior $\left( \frac{R}{2\rho _{0}}\right) ^{-2}\frac{1}{\sqrt{\pi }\Gamma
(-1/2)}\frac{\mathbf{d}_{1}\cdot \mathbf{d}_{2}}{\rho _{0}R}=-\frac{2\rho
_{0}}{\pi }\frac{\mathbf{d}_{1}\cdot \mathbf{d}_{2}}{R^{3}}$. While the
asymptotic of the second term is $\left( \frac{R}{2\rho _{0}}\right) ^{-3}%
\frac{1}{\sqrt{\pi }\Gamma (-3/2)}\frac{\mathbf{R\mathbf{\cdot }d_{1}R\cdot d%
}_{2}}{\rho _{0}^{2}R^{2}}=\frac{6\rho _{0}}{\pi }\frac{\mathbf{R\mathbf{%
\cdot }d_{1}R\cdot d}_{2}}{R^{5}}$. Combining the latter expressions we
obtain

\begin{equation}
V_{\text{dd}}\left( \mathbf{R}\right) \xrightarrow[r \rightarrow \infty]{}%
\frac{k}{\kappa }\frac{1}{R^{3}}\left[ \mathbf{d}_{1}\cdot \mathbf{d}_{2}-3%
\frac{\left( \mathbf{R\cdot d}_{1}\right) \left( \mathbf{R\cdot d}%
_{2}\right) }{R^{2}}\right] .  \label{DDRK}
\end{equation}%
For comparison the dipole-dipole interaction in 3D configuration space has
the following form

\begin{equation}
V_{\text{dd}}^{C}\left( \mathbf{r}\right) =\frac{k}{\kappa }\frac{1}{R^{3}}%
\left[ \mathbf{d}_{1}\cdot \mathbf{d}_{2}-3\frac{\left( \mathbf{R\cdot d}%
_{1}\right) \left( \mathbf{R\cdot d}_{2}\right) }{R^{2}}\right] ,
\label{DDC}
\end{equation}%
where $\kappa =\epsilon $ is the dielectric constant of the bulk material.
Thus, $V_{\text{dd}}\left( \mathbf{R}\right) $ asymptotic coincides with the
dipole-dipole interaction in 3D configuration space where charges interact
via the Coulomb potential.
\begin{figure}[h]
\noindent
\begin{centering}
\includegraphics[width=7.5cm]{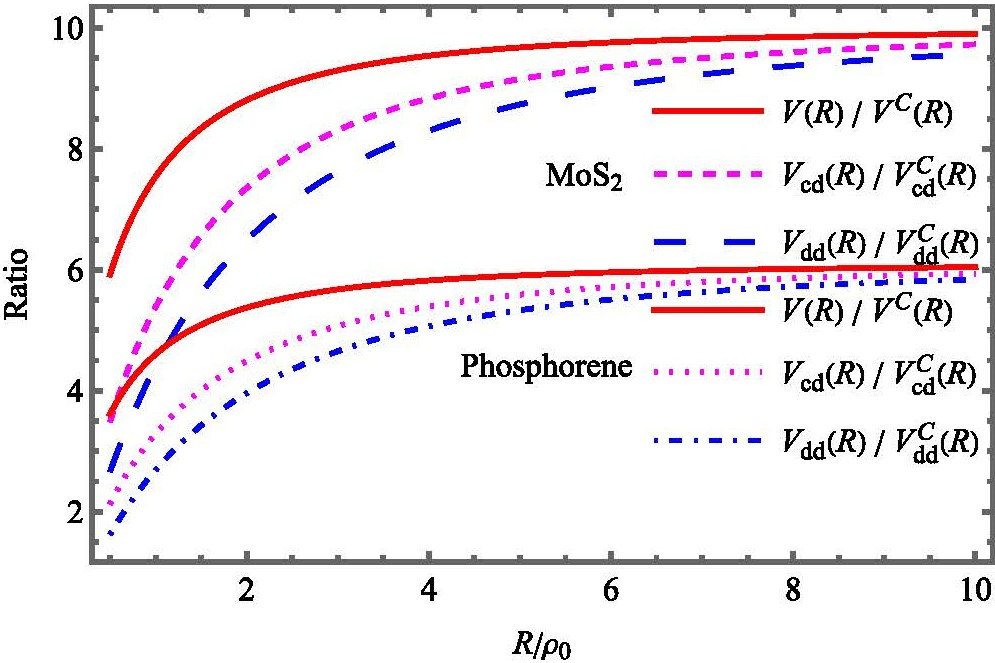}
\includegraphics[width=7.5cm]{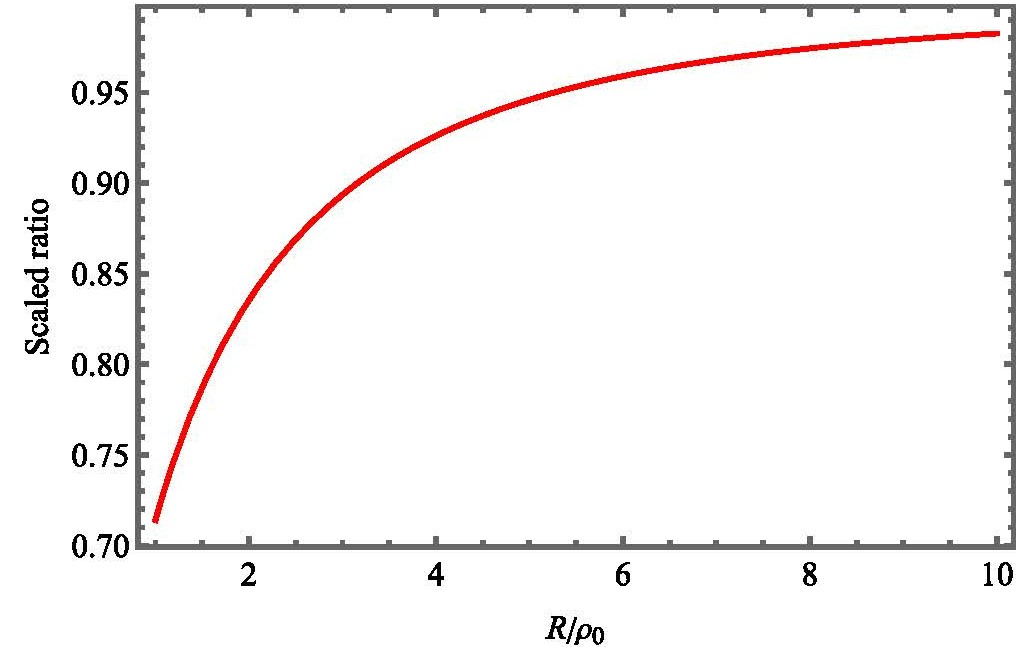}
\par\end{centering}
\caption{(Color online) Left panel: The ratios of the Rytova-Keldysh and
Coulomb potentials, charge-dipole interaction $V_{\text{cd}}$ in 2D
configuration space and $V_{\text{cd}}^{C}$ for bulk materials and the
second factors of the dipole-dipole interaction in a monolayer and bulk
material. Calculations are performed for the phosphorene and MoS$_{2}$.
Right panel: The universal dependence of the ratio of $V_{\text{cd}}$/$V_{%
\text{cd}}^{C}$ and $V(r)$/$V(R)^{C}$ on $R/\protect\rho_{0}$ for any 2D
material.}
\label{ChargeCharge}
\end{figure}
In numerical calculations, we focus only on freestanding phosphorene and
monolayer MoS$_{2}$. We use for MoS$_{2}$ polarizability $\chi _{2D}=6.6$
\AA\ \cite{Berkelbach2013} obtained within density functional theory and
subsequent the random phase approximation calculations for TMDC monolayers.
For the phosphorene polarizabilty the value $\chi _{2D}=4.1$ \AA\ \cite%
{Rodin2014} is used. The right panel in Fig. \ref{ChargeCharge} presents the
ratio of charge-dipole interaction potentials $V_{\text{cd}}\left( \mathbf{R}%
\right) $ in phosphorene and MoS$_{2}$ and $V_{\text{cd}}^{C}\left( \mathbf{R%
}\right) $ in the same bulk materials. The values of the negative order
Struve $H_{-1}\left( \frac{R}{\rho _{0}}\right) $ and Bessel $Y_{-1}\left(
\frac{R}{\rho _{0}}\right) $ functions were evaluated with the in-built
codes in Mathematica. There are five distinguished features: i. the value of
$V_{\text{cd}}\left( \mathbf{r}\right) \ $is bigger than the value of $V_{%
\text{cd}}^{C}\left( \mathbf{R}\right) $; ii. at small distances $V_{\text{cd%
}}\left( \mathbf{r}\right) $ falls more slowly than the Coulomb potential
induced charge-dipole interaction in the same bulk material; iii. at small
distances the slope of the ratio fall demonstrate the sensitivity of $V_{%
\text{cd}}\left( R\right) $ to the 2D polarizability and dependence on the
ratio of dielectric constant of the bulk material and the polarizability of
monolayer; v. the asymptotic of the ratio is the value of the dielectric
constant of the bulk material. This means that when $R\rightarrow \infty $
the charge-dipole interaction in a monolayer is the same as in vacuum.

The both dipole-dipole interactions \ref{DDK}) and (\ref{DDC}) have two
terms: one is proportional to $\mathbf{d}_{1}\cdot \mathbf{d}_{2}$ and the
other one to $\left( \mathbf{R\cdot d}_{1}\right) \left( \mathbf{R\cdot d}%
_{2}\right) $. The comparison of factors in front of $\mathbf{d}_{1}\cdot
\mathbf{d}_{2}$ shows that their ratio has the same dependence as the ratio $%
V_{\text{cd}}$/$V_{\text{cd}}^{C}$. The ratios of factors in front of $%
\left( \mathbf{R\cdot d}_{1}\right) \left( \mathbf{R\cdot d}_{2}\right) $
for phosphorene and MoS$_{2}$ are shown in Fig. \ref{ChargeCharge}. These
ratios are smaller than $V_{\text{cd}}$/$V_{\text{cd}}^{C}$ and demonstrate
the same features as that are listed above for $V_{\text{cd}}$/$V_{\text{cd}%
}^{C}$. However, the ratios fall more smoothly than $V_{\text{cd}}$/$V_{%
\text{cd}}^{C}$. As it is seen from Fig. \ref{ChargeCharge} $V$/$V^{C}>V_{%
\text{cd}}$/$V_{\text{cd}}^{C}>V_{\text{dd}}$/$V_{\text{dd}}^{C}$ and all
ratios converging to the dielectric constant of bulk materials. At small
distances $V$/$V^{C}$ increases more rapidly than $V_{\text{cd}}$/$V_{\text{%
cd}}^{C}$ and $V_{\text{cd}}$/$V_{\text{cd}}^{C}$ increases more fast than
the second term of $V_{\text{dd}}$/$V_{\text{dd}}^{C}$. Interestingly
enough, the ratio

\begin{equation}
\lbrack V_{\text{cd}}/V_{\text{cd}}^{C}]/\left[ V/V^{C}\right] =x\frac{%
H_{-1}(x)-Y_{-1}(x)}{H_{0}(x)-Y_{0}(x)},\text{ where \ }x=\frac{R}{\rho _{0}}
\end{equation}%
shows the universality in its dependence on $\frac{R}{\rho _{0}}$ that is
the same for any monolayer material. This ratio we named as a scaled ratio
that is the ratio of $V_{cd}(R)$ scaled to the corresponding $V_{cd}^{C}(R)$
and $V(R)$ scaled to the Coulomb potential. The dependence of this ratio on $%
R/\rho_{0}$ is shown on the right panel in Fig. \ref{ChargeCharge}.

\label{conc}

\textit{Concluding remarks.} In this paper we study the influence of the
reduction of dimensionality on the charge-dipole and dipole-dipole
interactions in 2D configuration space. We demonstrate that the screened
nature of Coulomb interaction imposes peculiarities in the 2D charge-dipole
and dipole-dipole interactions behavior. The analytical expression for the
charge-dipole and dipole-dipole interactions in 2D configuration space are
derived. We hope these charge-dipole and dipole-dipole interactions will find wide application in 2D materials studies. 
Recently, a new potential form for the electron-hole interaction,
which takes into account the three atomic sheets that compose a monolayer of
transition-metal dichalcogenides was derived \cite{Dery2018}. Without losing
any generality our approach can be extended for this form of the potential.

\end{document}